\documentstyle[12pt,epsf]{article}

\makeatletter
\@addtoreset{equation}{section}
\makeatother

\renewcommand{\today}{\ifcase\month\or
 Jan.\or Feb.\or Mar.\or Apr.\or May\or Jun.\or
 Jul.\or Aug.\or Sep.\or Oct.\or Nov.\or Dec.\fi
 \space\number\day, \number\year}

\newcommand{\mib}[1]{\mbox{\boldmath $#1$}}
\def\ie{{\it i.e.}}
\def\eg{{\it e.g.}}
\newcommand{\bra}[1]{\left\langle {#1}\right|}
\newcommand{\ket}[1]{\left|{#1}\right\rangle }
\newcommand{\VEV}[1]{\left\langle \,{#1}\,\right\rangle }
\newcommand{\tr}{\hbox{tr}}
\newcommand{\nn}{\nonumber\\}
\def\T{{\rm T}}
\def\r{{\rm r}}
\def\dmc{D_\mu c^a}
\def\dmcb{D_\mu \bar c^a}
\def\dncb{D_\nu \bar c}
\def\QB{Q_{\rm B}}
\def\qbdm{\{\QB,\,\dmcb\}}
\def\Qbdm{\big\{\QB,\,\dmcb\big\}}
\def\Qbdn{\big\{\QB,\,D_\nu \bar c^a\big\}}
\def\kugorel#1\over #2{\mathrel{\mathop{\kern0pt #1}\limits_{#2}}}
\def\llongrightarrow{\relbar\kern-3pt\longrightarrow}
\def\Green#1{\bra{0}\T\,#1\ket{0}}
\begin{document}

\begin{titlepage}

\begin{flushright}
\begin{minipage}[t]{3cm}
KUNS-1368 \\
HE(TH) 95/18\\
hep-th/9511033 \\
July, 1994
\end{minipage}
\end{flushright}

\vspace{0.7cm}

\renewcommand{\baselinestretch}{1.35}

\begin{center}
\Large\bf
The Universal Renormalization Factors \ $Z_1/Z_3$ \\
and Color Confinement Condition \\
in Non-Abelian Gauge Theory\footnote{%
Talk given at International Symposium on BRS Symmetry, Sept.~18 -- 22,
1995, Kyoto.}
\end{center}

\vfill

\begin{center}
{\large
Taichiro {\sc Kugo}\footnote{
E-mail address: kugo@gauge.scphys.kyoto-u.ac.jp
}}
\vskip 1.5em
{\it Department of Physics, Kyoto University, Kyoto 606-01, Japan}
\end{center}

\vfill

\vskip 2cm

\begin{abstract}
The ratio $Z_1/Z_3$ of
vertex and wave-function renormalization factors, which is universal
(\ie, matter-independent), is shown to equal $1+u$ which gives
the residue of the scalar pole $\propto  p_\mu p_\nu /p^2$ of 2-point function
$\VEV{D_\mu c\,\,\dncb}$. This relation is interesting since $1+u=0$ has
been known to give a sufficient condition for color confinement. We
also give an argument that, when $1+u=0$ holds, it will be
realized by the
disappearance of the massless gauge boson pole and is related with
the restoration of a certain ``local gauge symmetry" as was discussed
by Hata.
\end{abstract}
\end{titlepage}
\setcounter{footnote}{0}

\section{Introduction}

It is a well-known consequence of the Slavnov-Taylor identity that
the ratio of vertex renormalization factor
to wave-function renormalization factor
is {\it universal}:
\begin{equation}
{Z_1\over Z_3} = {\widetilde Z_1\over \widetilde Z_3}
= {Z_{\bar\psi \psi A}\over Z_{\psi }} = \ \cdots   \ ,
\label{eq:}
\end{equation}
where the denominators $Z_3$, $\widetilde Z_3$ and $Z_{\psi }$ are
the wave-function renormalization factors of
gauge-boson, Faddeev-Popov ghost and matter field $\psi $, respectively,
and the numerators $Z_1$, $\widetilde Z_1$ and
$Z_{\bar\psi \psi A}$ are the gauge-boson vertex renormalization factors of
those fields. Namely, the ratio is independent of
the measured matter fields (and equals 1 in the QED case as a result
of the Ward identity).
[We may, however, have to keep in mind that it is
{\it gauge-dependent}.]

The main purpose of this talk is to show
that the following equality holds for
this universal renormalization factor $Z_1/Z_3$ generally
in covariant gauges (with arbitrary gauge parameter $\alpha $):
\begin{equation}
{Z_1\over Z_3} = 1+u \ ,
\label{eq:MAIN}
\end{equation}
where $u\equiv  u(p^2\!=\!0)$ and $u(p^2)$ is the function defined by
\begin{equation}
\int d^4x\,e^{ipx}\bra{0}\T \dmc(x)\,\dncb^b(0)\ket{0}
= \left[\Bigl(g_{\mu \nu }-{p_\mu p_\nu \over p^2}\Bigr)
u(p^2)-{p_\mu p_\nu \over p^2}\right]\delta _{ab}\ .
\label{eq:DEFU}
\end{equation}
This relation (\ref{eq:MAIN}) is very interesting since
it is known\cite{ref:KO}
that $1+u=0$ gives a sufficient condition for all the
colored states to become unphysical; namely,
\begin{equation}
1+u=0 \quad  \Longrightarrow \quad \hbox{Color Confinement}\,!
\end{equation}

As a preparation for it, we briefly explain in Sect.~2 why the
condition $1+u=0$ implies the color confinement.
And then in Sect.~3 we give a proof of the relation (\ref{eq:MAIN}).
Some implications are discussed in Sect.~4, where I give an
argument that the confinement condition $1+u=0$ implies the
disappearance of the massless (vector) gauge boson pole. It,
therefore, turns to imply that a certain ``local gauge symmetry"
is dynamically restored as was discussed by Hata\cite{ref:Hata}.
This is explained in Sect.~5. The final Sect.~6 is devoted to
some further discussions.

\font\tenMIB = cmmib10

\section{ $1+u=0$: A Color Confinement Condition }

Let us first recapitulate how the condition $1+u=0$ is related to the
color confinement\cite{ref:KO}.
As noted by Ojima\cite{ref:Ojima} first, the equation of motion for
the gauge field can be written in the following form of Maxwell-type:
\begin{equation}
gJ_\mu ^a = \partial ^\nu F_{\mu \nu }^a + \qbdm \ ,
\label{eq:MAXWELL}
\end{equation}
where $J_\mu ^a$ is the Noether current of color symmetry (global gauge
symmetry). The point is that the Noether current always has an
{\it arbitrariness} adding a term of the form $\partial ^\nu f_{[\mu \nu ]}$
with an arbitrary local anti-symmetric tensor $f_{[\mu \nu ]}$.
That is, the modified current
$J_\mu ^{'a}\equiv  J_\mu ^a+\partial ^\nu f_{[\mu \nu ]}$ is still conserved
and, moreover,
the corresponding charge generates the infinitesimal color rotation
correctly on any field operators (at least with formal application of
canonical commutation relations).
We have, therefore, a possibility to define the color charge $Q^a$ by
\begin{eqnarray}
Q^a &=& \int d^3x\,\Bigl(J_\mu ^a - {1\over g}\partial ^\nu F_{\mu \nu
}^a\Bigr)_{\mu =0} \nn
    &=& \int d^3x\,{1\over g}\Qbdm_{\mu =0}
\label{eq:NULL}
\end{eqnarray}
If we {\it could} define the color charge by this equation, then
it takes a BRS-exact form and so the color confinement is concluded:
indeed, for any physical states $\ket{{\rm phy}}$ specified by
the condition $\QB\ket{{\rm phys}}=0$, we have
\begin{equation}
\bra{{\rm phys}}Q^a\ket{{\rm phys}'}=0 \ .
\end{equation}
It is an easy exercise to show from this equation that
{\it all the physical (that is, BRS-singlet) particles are
color-singlet}\cite{ref:KO}.

The expression Eq.~(\ref{eq:NULL}), however, does {\it not} give
a well-defined color charge operator generally. This is because
there is a massless one-particle contribution to
$J_\mu ^a$, $\partial ^\nu F_{\mu \nu }^a$ and $\qbdm$ so that the 3-volume
integration does not converge. To show this, we have first
to explain the elementary quartet (a quartet = a pair of BRS-doublets).

We can easily show that there always exists a massless quartet for
each group index $a$. Indeed, using
the equation of motion $\partial ^\mu \dmc=0$ and
equal time anti-commutation relation, we find an identity
\begin{equation}
\partial ^\mu _x\bra{0}\T \dmc(x)\, \bar c^b(y) \ket{0}
= \delta ^{ab}\delta ^4(x-y)\ ,
\end{equation}
implying that
\begin{equation}
\int d^4x\,e^{ipx}\bra{0}\T \dmc(x)\, \bar c^b(0) \ket{0}
= i\delta ^{ab}{p_\mu \over p^2} \ .
\label{eq:WT1}
\end{equation}
But, an identity
$\bra{0}\{ \QB, \T(A_\mu ^a\,\bar c^b)\}\ket{0}
=0$ means the Ward-Takahashi identity
\begin{equation}
i\bra{0}\T \dmc(x)\, \bar c^b(y) \ket{0}
= \bra{0}\T A_\mu ^a(x)\, B^b(y) \ket{0} \ ,
\end{equation}
so that Eq.~(\ref{eq:WT1}) implies also an identity:
\begin{equation}
\int d^4x\,e^{ipx}\bra{0}\T A_\mu ^a(x)\, B^b(0) \ket{0}
= -\delta ^{ab}{p_\mu \over p^2} \ .
\label{eq:WT2}
\end{equation}
The identities (\ref{eq:WT1}) and (\ref{eq:WT2}) give exact Green
functions and so the presence of massless pole $\propto   1/p^2$
implies that the following four massless asymptotic fields
$\gamma ^a$, $\bar\gamma ^a$, $\chi ^a$ and $\beta ^a$ exist for each group
index $a$:
\begin{eqnarray}
\dmc(x)
&\kugorel\llongrightarrow \over {x_0\rightarrow  \pm  \infty   }&
\partial _\mu \gamma ^a(x) + \cdots  \nn
\bar c^a(x) & \longrightarrow& \bar \gamma ^a(x) + \cdots   \nn
A_\mu ^a(x) & \longrightarrow& \partial _\mu \chi ^a(x) + \cdots   \nn
B^a(x) & \longrightarrow& \beta ^a(x) + \cdots   \ .
\label{eq:QUARTET}
\end{eqnarray}
{}From BRS transformation law $[i\QB,\,A_\mu ^a]=\dmc$
and $\{\QB,\,\bar c^a\}=B^a$, we find that these asymptotic
fields transform as follows under the BRS transformation:
\begin{equation}
[i\QB,\,\chi ^a(x)]=\gamma ^a(x)\ , \qquad
\{\QB,\,\bar \gamma ^a(x)\}=\beta ^a(x) \ .
\end{equation}
This clearly shows that these asymptotic fields
$\gamma ^a$, $\bar\gamma ^a$, $\chi ^a$ and $\beta ^a$ really forms a
BRS quartet representation.

Now we come back to the well-definedness problem of the expression
Eq.~(\ref{eq:NULL}). The point is that
$\dmcb=\partial _\mu \bar c^a + g(A_\mu \times  \bar c)^a$ generally contains
the one-particle contribution of the massless asymptotic field
$\bar \gamma ^a$:
\begin{equation}
\dmcb=\partial _\mu \bar c^a + g(A_\mu \times  \bar c)^a
\quad  \longrightarrow \quad  (1+u)\partial _\mu \bar\gamma ^a \ ,
\end{equation}
where the weight 1 comes from the first term
$\partial _\mu \bar c^a$ and the weight $u$ from the second
$g(A_\mu \times  \bar c)^a$. This can be seen by looking at the definition
(\ref{eq:DEFU}) of $u$ and Eq.~(\ref{eq:WT1}). We thus
see that the operator $\qbdm$ contains the massless
one-particle contribution of
the elementary quartet member $\beta ^a$ in the form
\begin{equation}
\big\{\,\QB,\,\dmcb(x)\,\big\}
\quad \kugorel\llongrightarrow \over {x_0\rightarrow  \pm  \infty   }
\quad (1+u)\partial _\mu \beta ^a(x) \ .
\end{equation}
Generally, the other operators $J_\mu ^a$ and
$\partial ^\nu F_{\mu \nu }^a$ appearing in the Maxwell equation
(\ref{eq:MAXWELL})
also have the one-particle contributions
from $\beta ^a$ since they carry the same quantum numbers as
$\qbdm$:
\begin{equation}
J_\mu ^a(x) \quad \longrightarrow\quad  v \partial _\mu \beta ^a(x) \ ,
\qquad \partial ^\nu F_{\mu \nu }^a(x) \quad \longrightarrow\quad -w \partial
_\mu \beta ^a(x) \ ,
\end{equation}
with suitable weights $v$ and $w$. In QED, one can easily see that
$u=0$ and  $w=Z_3$. The Maxwell equation (\ref{eq:MAXWELL})
tells us the relation between those weights:
\begin{equation}
gv=-w+(1+u)\ .
\label{eq:KANKEI}
\end{equation}
The well-defined color charge is, therefore, given by
\begin{equation}
Q^a = \int d^3x\,\Bigl(J_\mu ^a
+ {v\over w}\partial ^\nu F_{\mu \nu }^a\Bigr)_{\mu =0} \ ,
\label{eq:WELLDEFINED}
\end{equation}
so that the massless one-particle contribution $\partial _\mu \beta ^a$
contained in $J_\mu ^a$ is cancelled by that in $\partial ^\nu F_{\mu \nu }^a$.

Now, if the condition $1+u=0$ holds, then we have
$v/w=-1/g$ and so the well-defined color charge
Eq.~(\ref{eq:WELLDEFINED}) happens to coincide with
the previous BRS-exact expression (\ref{eq:NULL}) implying the
color confinement.

\section{Proof of the relation $Z_1/Z_3=1+u$}

I have noticed this relation $Z_1/Z_3=1+u$ for the first time in
studying the renormalization problem in the background field method
in collaboration with Imamura and Van Proeyen\cite{ref:IKP}.
However, the proof by that method is a bit complicated and so here
I will give a simpler proof using a similar method to
Aoki's\cite{ref:Aoki} which
he used when proving the electro-magnetic charge universality in the
Weinberg-Salam model.

Add $g(v/w)\partial ^\nu F_{\mu \nu }^a$ to both sides of the Maxwell equation
(\ref{eq:MAXWELL}), and then we have
\begin{equation}
gJ_\mu ^{a,\rm well\hbox{-}def} =
\Bigl(1+g{v\over w}\Bigr)\partial ^\nu F_{\mu \nu }^a + \qbdm \ ,
\end{equation}
with
\begin{equation}
J_\mu ^{a,\rm well\hbox{-}def} \equiv   J_\mu ^a
+ {v\over w}\partial ^\nu F_{\mu \nu }^a
\end{equation}
being the color current giving the well-defined color charge
(\ref{eq:WELLDEFINED}). Using a relation $1+gv/w=(1+u)/w$
following from Eq.~(\ref{eq:KANKEI}), we find
\begin{equation}
{1+u\over w}\partial ^\nu F_{\mu \nu }^a =
gJ_\mu ^{a,\rm well\hbox{-}def} -  \qbdm \ .
\end{equation}
Sandwitching this with two physical states $\bra{f}$ and $\ket{g}$
satisfying $\bra{f}\QB=\QB\ket{g}=0$, yields
\begin{equation}
{1+u\over w}\partial ^\nu \bra{f}F_{\mu \nu }^a(x)\ket{g} =
g\bra{f}J_\mu ^{a,\rm well\hbox{-}def}(x)\ket{g} \ .
\label{eq:AOKI}
\end{equation}

We now make
an operation $\lim_{p\rightarrow  0}\int d^4x\,e^{ipx}$
act on both sides of this equation and evaluate the both sides
separately. Start with the left-hand
side. Because of the presence of the
divergence $\partial ^\nu $, only massless one-particle
pole contribution to the matrix element
$\bra{f}F_{\mu \nu }^a(x)\ket{g}$ can survive in this limit. But such a
massless one-particle pole (if any) comes from the gauge boson
contribution as shown in Fig.~1
\begin{figure}
\vskip .2cm
\epsfxsize=4.5cm
\centerline{\epsffile{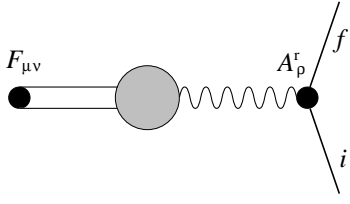}}
\caption{Diagram of massless pole contribution.}
\end{figure}
and generally takes the form
\begin{eqnarray}
&&\int d^4x\,e^{ipx}
\bra{f}F_{\mu \nu }^a(x)\ket{g}\Big|_{\rm pole\ part}
=\VEV{F_{\mu \nu }^a\,A^b_{\r,\rho }} V^{\rho ,b}_{fi}\ , \nn
&&\quad \VEV{F_{\mu \nu }^a\,A^b_{\r,\rho }}
= -Y{p_\mu g_{\nu \rho }-p_\nu g_{\mu \rho }\over p^2} \delta _{ab} \ ,\nn
&&\quad V^{\rho ,b}_{fi}=
ig_\r T^b_{fi} (p_f+p_i)^\rho  (2\pi )^4\delta ^4(p+p_f-p_i)\ ,
\label{eq:POLE}
\end{eqnarray}
where $\VEV{F_{\mu \nu }^a\,A^b_{\r,\rho }}$ stands for the propagator from
$F_{\mu \nu }^a$ to the renormalized gauge boson $A^b_{\r,\rho }$, and
$V^{\rho ,b}_{fi}$ for the vertex of the
renormalized gauge boson between the
on-shell initial and final states. $Y$ is a weight factor and
$g_\r$ is the renormalized coupling constant.
Substituting this expression, we find that
the left-hand side becomes
\begin{equation}
-Y{1+u\over w}\lim_{p\rightarrow  0}\Bigl(-i{p_\mu p_\rho \over p^2}+ig_{\mu
\rho }\Bigr)
(ig_\r)T^a_{fi}(p_f+p_i)^\rho  (2\pi )^4\delta ^4(p+p_f-p_i)\ .
\end{equation}
Noting the pole term $p_\mu p_\rho /p^2$ vanishes owing
to the ``current conservation" at the vertex, we obtain
\begin{eqnarray}
&&{1+u\over w}\lim_{p\rightarrow  0}\int d^4x\,e^{ipx}
\partial ^\nu \bra{f}F_{\mu \nu }^a(x)\ket{g} \nn
&& \qquad \qquad \qquad \ = \
Y{1+u\over w}g_\r T^a_{fi}\lim_{p_f\rightarrow  p_i}
(p_f+p_i)_\mu  (2\pi )^4\delta ^4(p_f-p_i)\ .
\label{eq:LHS}
\end{eqnarray}
The weight factor $Y$ introduced in Eq.~(\ref{eq:POLE}) can easily
be related to the previous weight $w$: using
$\partial ^\nu F_{\mu \nu }^a(x) \longrightarrow -w \partial _\mu \beta ^a(x)$
and the WT identity
(\ref{eq:WT2}) with $A_{\r,\rho }^b=Z_3^{-1/2}A_\rho ^b$, we obtain
\begin{equation}
\VEV{\partial ^\nu F_{\mu \nu }^a\,A_{\r,\rho }^b}\Big|_{\rm pole\ part} =
-w \VEV{\partial _\mu B^a\,A_{\r,\rho }^b}
= w Z_3^{-1/2}(i\delta ^{ab}){p_\mu p_\rho \over p^2} \ ,
\end{equation}
comparison of which with the divergence of the second equation
in Eq.~(\ref{eq:POLE}) leads to
\begin{equation}
Y=w Z_3^{-1/2}\ , \qquad   \hbox{so that} \qquad
Y{1+u\over w}= (1+u)Z_3^{-1/2}.
\label{eq:KANKEI2}
\end{equation}

The right hand side of Eq.~(\ref{eq:AOKI}),
on the other hand, can be easily evaluated for $\mu =0$ component
as follows:
\begin{eqnarray}
\lim_{p\rightarrow  0}\int d^4x\,e^{ipx}
g\bra{f}J_0^{a,\rm well\hbox{-}def}(x)\ket{g} &=&
\int dx_0\,g\bra{f}Q^{a,\rm well\hbox{-}def}\ket{g} \nn
&=&
g T^a_{fi} \lim_{p_f\rightarrow  p_i}2p_i^0(2\pi )^4\delta ^4(p_f-p_i) \ .
\end{eqnarray}
Now comparing this with the above Eq.~(\ref{eq:LHS}) with
Eq.~(\ref{eq:KANKEI2}), we finally
obtain a relation
\begin{equation}
(1+u)Z_3^{-1/2}g_\r = g \ .
\end{equation}
But, if we use the relation $g=Z_1Z_3^{-3/2}g_\r$
between the bare
and renormalized coupling constants $g$ and $g_\r$, this is nothing
but the desired relation $1+u=Z_1/Z_3$, Eq.~(\ref{eq:MAIN}).

We have proved the relation $1+u=Z_1/Z_3$ in covariant gauges with
arbitrary values of gauge parameter $\alpha $. In Landau gauge $\alpha =0$,
however, it is easier to show it.
First define the following two-point functions
\begin{eqnarray}
\VEV{c\,\,\bar c} &\equiv  & -{1\over p^2G(p^2)}\ , \nn
\VEV{g(A_\mu \!\!\times  \!c)\,\,\bar c}_{\rm 1PI} &\equiv  &
\VEV{g(A_\mu \!\!\times  \!c)\,\,\bar c}/\VEV{c\,\,\bar c}
\equiv   -ip_\mu F(p^2)\ .
\end{eqnarray}
Here and henceforth we use an abbreviated notation:
\begin{equation}
\VEV{\,A\,B\,} \equiv   \int d^4x\,e^{ipx}\Green{A(x)\,B(0)}
\ .
\end{equation}
The suffix 1PI generally denotes the one-particle irreducible vertex.
Then, we have
\begin{eqnarray}
\VEV{D_\mu c\,\,\bar c}
= \VEV{\partial _\mu c\,\,\bar c}+\VEV{g(A_\mu \!\!\times  \!c)\,\,\bar c}
\equiv   ip_\mu \big(1+F(p^2)\big){1\over p^2G(p^2)} \ .
\end{eqnarray}
But this must equal $ip^\mu /p^2$ because of Eq.~(\ref{eq:WT2})
so that
\begin{equation}
1+F(p^2)= G(p^2) \ .
\end{equation}
On the other hand,
a simple diagramatical consideration shows that
\begin{equation}
\VEV{c\,\,g(A_\nu \!\!\times  \!\bar c)}_{\rm 1PI} =
ip^\rho \VEV{g(A_\rho \!\!\times  \!c)\,\,g(A_\nu \!\!\times  \!\bar c)}_{\rm
1PI}
\end{equation}
and hence that
\begin{eqnarray}
\VEV{D_\mu c\,\,g(A_\nu \!\!\times  \!\bar c)}
&=&\VEV{D_\mu c\,\,\bar c}\,\VEV{c\,\,g(A_\nu \!\!\times  \!\bar c)}_{\rm 1PI}
+ \VEV{g(A_\mu \!\!\times  \!c)\,\,g(A_\nu \!\!\times  \!\bar c)}_{\rm 1PI}
\nn
&=&\Bigl(\delta _\mu ^\rho -{p_\mu p^\rho \over p^2}\Bigr)
\VEV{g(A_\rho \!\!\times  \!c)\,\,g(A_\nu \!\!\times  \!\bar c)}_{\rm 1PI}\ .
\end{eqnarray}
The last Green function
$\VEV{D_\mu c\,\,g(A_\nu \!\!\times  \!\bar c)}$ must be equal to
$\big(g_{\mu \nu }-(p_\mu p_\nu /p^2)\big)u(p^2)$ because of the
definition (\ref{eq:DEFU}) of the function $u(p^2)$ so that
$\VEV{g(A_\mu \!\!\times  \!c)\,\,g(A_\nu \!\!\times  \!\bar c)}_{\rm 1PI}$
have to have the form
\begin{equation}
\VEV{g(A_\mu \!\!\times  \!c)\,\,g(A_\nu \!\!\times  \!\bar c)}_{\rm 1PI}
= g_{\mu \nu }u(p^2)+p_\mu p_\nu v(p^2)
\label{eq:ACAB}
\end{equation}
with $v(p^2)$ being an arbitrary function. Up to here all the
equations hold for any covariant $\alpha $ gauges. But, in Landau
gauge $\alpha =0$, we have additionally an identity
\begin{eqnarray}
-ip_\mu F(p^2) &\equiv  &
\VEV{g(A_\mu \!\!\times  \!c)\,\,\bar c}_{\rm 1PI} \nn
&\equiv  &
\VEV{g(A_\mu \!\!\times  \!c)\,\,g(A_\nu \!\!\times  \!\bar c)}_{\rm 1PI}
\times  (-ip^\nu ) \ .
\end{eqnarray}
This is because the derivative factor acting on the anti-ghost at
the $\bar c$-$A_\nu $-$c$ vertex at the right end of the diagrams
can be transferred to act on the external ghost
since $\partial _\nu A^\nu =0$ in the Landau gauge. Therefore we have
\begin{equation}
-ip_\mu F(p^2)= -ip_\mu \Bigl(u(p^2)+p^2v(p^2)\Bigr)\ ,
\quad \Longrightarrow \quad
G(p^2) -1 = u(p^2)+p^2v(p^2)\ .
\end{equation}
This relation yields at $p^2=0$
\begin{equation}
G(0) = 1 + u(0) \ ,
\end{equation}
but, since $G(0)=\widetilde Z_3^{-1}$ and $\widetilde Z_1 = 1$
in the Landau gauge, this shows the relation
$\widetilde Z_1/\widetilde Z_3 = 1+u$, thus proving the desired
identity Eq.~(\ref{eq:MAIN}).

\section{What Does $1+u=Z_1/Z_3$ Imply?}

How is $1+u=Z_1/Z_3=0$ expected to be realized?
Naively, we immediately expect the following two possibilities:
\newline\hspace*{1cm}
1. \ $Z_1=0$, and $Z_3={}$finite.
\newline\hspace*{1cm}
2. \ $Z_1={}$finite, and $Z_3=0$.
\newline
However, since the condition is concerned with only the ratio,
we have, for instance, even the possibility
\newline\hspace*{1cm}
3. \ $Z_3=0$, but/and $Z_1/Z_3=0$.
\newline
Here I mean that $Z_1$ has a higher zero than $Z_3$; \eg,
$Z_3(p^2)={\cal O}(p^2)$ and $Z_1(p^2)={\cal O}(p^4)$.
The wave-function renormalization factor $Z_3(p^2)$ as a
function of $p^2$ can be defined as follows by the general form of
gauge boson 2-point function:
\begin{equation}
\int d^4x\,e^{ipx}\Green{A_\mu ^a(x)\,A_\nu ^b(x)}
=\left[\Bigl(g_{\mu \nu }-{p_\mu p_\nu \over p^2}\Bigr)
{Z_3(p^2)\over p^2}+\alpha {p_\mu p_\nu \over p^4}\right]\delta ^{ab}
\end{equation}
Although I cannot prove it, it seems plausible that the possibility 3
is realized when $1+u=0$. Let us now explain this.

Assume that $Z_3(p^2\!=\!0)\not=0$, then, there exists a {\it vector}
asymptotic field ${\cal A}^a_\mu (x)$ in the channel $A_\mu ^a(x)$:
\begin{equation}
A_\mu ^a(x) \quad \kugorel\llongrightarrow \over {x_0\rightarrow  \pm  \infty
}
\quad  {\cal A}^a_\mu (x) \ .
\end{equation}
This is a colored particle. But any colored particles are confined
when $1+u=0$. Therefore the BRS transform of $A_\mu ^a(x)$
\begin{equation}
[i\QB, \, A_\mu ^a(x)] = \dmc(x) \ ,
\end{equation}
with ghost number 1, should also have its own massless vector
asymptotic field ${\cal C}^a_\mu (x)$ such that
$\big({\cal A}^a_\mu (x), \,{\cal C}^a_\mu (x)\big)$
form a BRS-doublet (and hence become unphysical):
\begin{equation}
[i\QB, \, {\cal A}_\mu ^a(x)] = {\cal C}^a_\mu (x) \ .
\end{equation}
This is a color confinement by the quartet mechanism\cite{ref:KUGO}.
Since this asymptotic field ${\cal C}^a_\mu (x)$ is a vector,
it must be a bound-state in the $(A_\mu \times  c)^a$ channel. So it is
natural\footnote{
This is the point which is not quite rigorous in this argument.
There remains a possibility that the vector
asymptotic field ${\cal C}^a_\mu (x)$ exists
but nevertheless does not produce a pole in the 1PI Green function
$\VEV{g(A_\mu \!\!\times  \!c)\,\,g(A_\nu \!\!\times  \!\bar c)}_{\rm 1PI}$.
This might not be so strange a possibility since we are now considering
in any case such an unfamiliar situation that
the always existing quartet
member $\beta ^a$ disappears from the operator $\qbdm$.
This point was emphasized by Izawa\cite{ref:IZAWA}.
But, if
$\VEV{g(A_\mu \!\!\times  \!c)\,\,g(A_\nu \!\!\times  \!\bar c)}_{\rm 1PI}$
does not show up the massless vector pole, then, where will the
asymptotic vector field ${\cal C}^a_\mu (x)$ produce its pole
at all?
}
to suppose that it should show up as a pole
in the 1PI Green function
$\VEV{g(A_\mu \!\!\times  \!c)\,\,g(A_\nu \!\!\times  \!\bar c)}_{\rm 1PI}$,
namely, in Eq.~(\ref{eq:ACAB})
$$
\VEV{g(A_\mu \!\!\times  \!c)\,\,g(A_\nu \!\!\times  \!\bar c)}_{\rm 1PI}
= g_{\mu \nu }u(p^2)+p_\mu p_\nu v(p^2) \ .
$$
The vector massless pole $1/p^2$ should appear
in the $g_{\mu \nu }$-part, \ie, in
$u(p^2)$. But this contradicts with the present assumption
$u=u(p^2\!=\!0)=1$.

This argument, therefore, suggests that the massless vector pole
in $A_\mu ^a$, which existed in the perturbation phase,
should disappear (or the appearance of mass gap): namely,
\begin{equation}
Z_3(p^2) \propto   p^2 \ ,
\end{equation}
corresponding to the above possibility 3.

\section{Dynamical Restoration of a ``Local" Gauge Symmetry}

The conclusion in the previous section reminds me Hata's
work\cite{ref:Hata,ref:Hata2} who clarified a symmetry aspect of
the `strange' condition $1+u=0$. So let us briefly recapitulate
what he has done in this context.

In QED, the covariant $\alpha $ gauge fixing still leaves a symmetry
under the gauge transformation with transformation parameter
linear in $x$:
\begin{equation}
\Lambda (x) = a_\nu x^\nu +b \ .
\end{equation}
Although being a (special) ``local" gauge symmetry,
this transformation can be regarded as a combination
of global transformations with $x$-independent parameters $a_\nu $
and $b$. We have accordingly conserved Noether charges,
vector one $Q^\nu $ and scalar one $Q$, (the latter $Q$ being
just a usual electro-magnetic (global $U(1)$) charge),
and they give the generators of the original transformation:
\begin{eqnarray}
&&[\,i(a_\nu Q^\nu +bQ), \,A_\mu (x)\,] = \partial _\mu \Lambda (x) = a_\mu \ ,
\nn
&& \qquad \qquad \Longrightarrow \quad
\cases{
[\,iQ, \,A_\mu (x)\,] = 0  \cr
[\,iQ^\nu , \,A_\mu (x)\,] = \delta _\mu ^\nu  \cr}
\end{eqnarray}
The first commutation relation says only that the photon $A_\mu $
carries vanishing charge, but the second one shows that the vector
symmetry of charge $Q^\nu $ is {\it always} spontaneously broken
since the right-hand side is non-vanishing c-number $\delta _\mu ^\nu $\,!
It also says that the massless
Nambu-Goldstone (NG) mode corresponding to the
spontaneous breaking should appear in the $A_\mu $ channel.
It can be argued that,
as far as the global $U(1)$ symmetry corresponding to the
scalar charge $Q$ is not spontaneously broken,
the NG mode must be a vector particle so that the photon can be
identified with the Nambu-Goldstone vector\cite{ref:FPBN}!
[This type of argument is important since it {\it proves} the
existence and exact masslessness of the photon. Quite a parallel
argument can apply to gravity\cite{ref:NO,ref:KTU} and proves the
existence and exact masslessness of the graviton.]

Hata considered the same thing in non-Abelian gauge theory.
Corresponding to the ``local" gauge transformation,
\begin{equation}
\Lambda ^a(x) = a_\nu ^ax^\nu +b^a \ ,
\label{eq:NATR}
\end{equation}
there appear the following scalar and vector charges,
both of which carry color index $a$ now:
\begin{eqnarray}
Q^a \quad  &\longrightarrow& \quad J_\mu ^a(x) \ , \nn
Q^{a,\nu } \quad  &\longrightarrow& \quad
J_\mu ^{a,\nu }(x)  = gJ_\mu ^a(x)x^\nu  + F_\mu ^{a\ \nu }\ .
\end{eqnarray}
In this case, however, the vector transformation with parameter
$a_\nu ^a$ is not an exact symmetry but is slightly violated and so
\begin{equation}
\partial ^\mu J_{\mu ,\nu }^a(x) = \Qbdn \ .
\end{equation}
In his second paper\cite{ref:Hata2}, Hata considered a modified
vector transformation which becomes an exact symmetry.
But here we continue
to consider the present simplified version. Again, this
``local symmetry" with parameter $a_\nu ^ax^\nu $ is spontaneously broken
in the perturbation phase and the usual perturb-ative massless
gauge boson can be identified with Nambu-Goldstone vector.

Contrary to the QED case, however, there is a possibility that
this ``local symmetry" is restored dynamically in the non-Abelian case.
Hata's picture for the color confinement is as follows:
\begin{eqnarray}
&&\hbox{Dynamical restoration of the ``local symmetry" with
parameter $a_\nu ^ax^\nu $} \nn
&& \quad = \quad \hbox{Disordered phase}
\quad = \quad \hbox{Confinement phase} \ .
\end{eqnarray}
Indeed, if the ``local symmetry" is dynamically restored, then
the massless vector pole (\ie, perturbative gauge boson pole) in
the current $J^a_{\mu ,\nu }$ should disappear and it will have no
massless pole contribution. This requirement actually leads to
the above condition $1+u=0$ of ours, since we have
\begin{equation}
\partial ^\mu \VEV{\,J_{\mu ,\nu }^a(x)\,\,A_\rho ^b\,}
= i\Bigl(g_{\mu \rho }-{p_\mu p_\rho \over p^2}\Bigr)\bigl(1+u(p^2)\bigr)
\ .
\end{equation}
Namely our condition $1+u=0$ is a necessary condition for the
restoration of the ``local symmetry". I do not know whether it is also
sufficient, but
our discussion in the previous section strongly suggests that
it {\it is}. Hata also gave another interesting direct proof that
any colored particle has to be a BRS-doublet if the current
$J_{\mu ,\nu }^a(x)$ has no massless pole, by considering a
3-point Green function
of $J_{\mu ,\nu }^a(x)$ and the two fields corresponding to that
colored particle\cite{ref:Hata}.

\section{Discussions}

One may wonder how it is possible at all to satisfy our
confinement condition $1+u=0$. This question is quite natural if
we note the fact that $u=u(p^2\!=\!0)$ is
both infrared and ultraviolet divergent in perturbation theory!

However, we should recall that quite a similar thing is actually
realized in the non-linear sigma models in two dimension.
As an example, consider the following $O(N)/O(N-1)$ sigma model
again following Hata\cite{ref:Hata}:
\begin{equation}
{\cal L}= {1\over 2} {(\partial \mib\phi )^2
\over \Bigl(1+{\displaystyle {\lambda \over 4}}\mib\phi ^2\Bigr)^2 }\ ,
\qquad \qquad
\mib\phi \equiv   \pmatrix{\phi _1 \cr
\vdots \cr \phi _{N-1} \cr}\ .
\end{equation}
The non-linearly realized $O(N)/O(N-1)$ symmetry is the
transformation:
\begin{equation}
\delta _i\phi _j = \Bigl(1-{\lambda \over 4}\mib\phi ^2\Bigr)\delta _{ij}
+ {\lambda \over 2}\phi _i\phi _j \ ,
\end{equation}
with $i=1,\ 2,\ \cdots  ,\ N-1$. In perturbation phase, this non-linear
symmetry is (spontaneously) broken and the fields $\phi _i$ are the
corresponding massless NG bosons. The restoration condition for
this symmetry is given by, in the leading order in $1/N$ expansion,
\begin{equation}
\VEV{\delta _i\phi _j} = \delta _{ij}
\Bigl(1-{\lambda \over 4}\left\langle  {\mib\phi ^2}\right\rangle  \Bigr) = 0\
{}.
\end{equation}
This is quite analogous to our condition $1+u=0$.
The vacuum expectation value
$-(\lambda /4)\left\langle  {\mib\phi ^2}\right\rangle  $
corresponds to our $u$, and is indeed
both infrared and ultraviolet divergent in perturbation
calculation! Nevertheless
we know\cite{ref:BLS}
that $\bigl(1-\lambda \left\langle  {\mib\phi ^2}\right\rangle  /4\bigr)$
is actually realized in this model.
It is known rather generally that a mass gap appears and
the nonlinear symmetry is dynamically restored in
a wide variety of two dimensional non-linear sigma models.

As a matter of fact, this resemblance of dynamical
symmetry restorations between Yang-Mills gauge theory in four
dimension and non-linear sigma model in two dimension, is not
a mere analogy.  Indeed, it becomes an exact correspondence if we
consider a pure gauge model of Yang-Mills theory
in four dimension\cite{ref:HK}.
Let us close this talk by briefly explaining this.

Take the $SU(N)$ Yang-Mills theory with
$OSp(4|2)$ symmetric gauge fixing:
\begin{equation}
{\cal L}= -{1\over 4}F_{\mu \nu }^2 +
\delta _{\rm B}\bar\delta _{\rm B}\Bigl({1\over 2}A_\mu ^2+i\bar cc\Bigr)
\end{equation}
where $\delta _{\rm B}$ and $\bar\delta _{\rm B}$ denote BRS and anti-BRS
transformations, respectively. The pure gauge model is given by
replacing the gauge field $A_\mu ^a$ in this Lagrangian by the
pure gauge mode $g^\dagger  \partial _\mu g$:
\begin{equation}
A_\mu (x) \qquad \longrightarrow \qquad  g^\dagger  (x)\partial _\mu g(x) \ .
\end{equation}
Then, the $F_{\mu \nu }^2$ term drops out and only the gauge-fixing
term remains. Namely the pure gauge model is a kind of topological
model. One can show very easily that this model is {\it exactly}
equivalent by the Parisi-Sourlas mechanism to the two dimensional
$SU(N)_{\rm L}\times  SU(N)_{\rm R}/SU(N)$ non-linear sigma model with
the action
\begin{equation}
\int d^2x\, \tr\bigl(\partial ^\mu g^\dagger \partial _\mu g\big) \ .
\end{equation}
This model has an $SU(N)_{\rm L}\times  SU(N)_{\rm R}$ symmetry
under the transformation $g \rightarrow   h_{\rm L}^\dagger  g h_{\rm R}$.
Very interestingly, the current corresponding to the
$SU(N)_{\rm R}$ symmetry is found to be given by
\begin{equation}
J_{{\rm R}\,\mu }^a = \Qbdm \ .
\end{equation}
So, if the symmetry is restored, then it implies that our
condition $1+u=0$ is literally realized. This is what actually occurs
as we know from the exact result due to
Polyakov and Wiegmann\cite{ref:PW}. So this pure gauge model
has a possibility to be used as a ``zeroth order" theory
in a new form of ``perturbation theory" in the actual non-Abelian
gauge theory, in which the confinement condition $1+u=0$ is
satisfied order by order. It is encouraging that such trials towards
this direction are already performed by several
authors\cite{ref:AN,ref:I,ref:HT}.

\section*{Acknowledgements}

I would like to thank H.~Hata, K.~Izawa and Y.~Taniguchi
for valuable discussions and comments.
He is supported in part by the Grant-in-Aid for
Cooperative Research (\#07304029) and the Grant-in-Aid for
 Scientific Research (\#06640387) from the Ministry of Education,
Science and Culture.


\end{document}